# Interrogating the superconductor $Ca_{10}(Pt_4As_8)(Fe_{2-x}Pt_xAs_2)_5$ Layer-by-layer

Jisun Kim[1], Hyoungdo Nam[2], Guorong Li[1], A. B. Karki[1], Zhen Wang[1,3], Yimei Zhu[3], Chih-Kang Shih[2], Jiandi Zhang[1], Rongying Jin[1], and E. W. Plummer[1,*]

[1]Department of Physics and Astronomy, Louisiana State University, Baton Rouge, LA 70803

[2]Department of Physics, The University of Texas, Austin, TX 78712

[3]Brookhaven National Laboratory, Upton, NY 11973

*Corresponding Author: E. W. Plummer (email: wplummer@phys.lsu.edu)

## Abstract

Ever since the discovery of high-$T_c$ superconductivity in layered cuprates, the roles that individual layers play have been debated, due to difficulty in layer-by-layer characterization. While there is similar challenge in many Fe-based layered superconductors, the newly-discovered $Ca_{10}(Pt_4As_8)(Fe_2As_2)_5$ provides opportunities to explore superconductivity layer by layer, because it contains both superconducting building blocks ($Fe_2As_2$ layers) and intermediate $Pt_4As_8$ layers. Cleaving a single crystal under ultra-high vacuum results in multiple terminations: an ordered $Pt_4As_8$ layer, two reconstructed Ca layers on the top of a $Pt_4As_8$ layer, and disordered Ca layer on the top of $Fe_2As_2$ layer. The electronic properties of individual layers are studied using scanning tunneling microscopy/spectroscopy (STM/S), which reveals different spectra for each surface. Remarkably superconducting coherence peaks are seen only on the ordered $Ca/Pt_4As_8$ layer. Our results indicate that an ordered structure with proper charge balance is required in order to preserve superconductivity.

Since the discovery of Fe-based superconductors in 2008, tremendous effort has been expended to understand the origin of their physical properties, that exhibit strong coupling between structure, magnetism, and superconductivity. Similar to high-temperature ($T_c$) cuprate superconductors, Fe pnictide superconductors form a layered structure with $Fe_2As_2$ building blocks. Depending on the spacers that separate these building blocks, these superconductors are often categorized as "111" (e.g. LiFeAs)[1], "1111" (e.g. LaFeAsO and CaFeAsF)[2, 3], and "122" (e.g. $BaFe_2As_2$)[4] families etc. Surprisingly, these compounds exhibit only small anisotropy in their physical properties, in spite of their layered structure[5, 6, 7, 8, 9]. This suggests strong $Fe_2As_2$ interlayer coupling. Cleaving a single crystal of such a system usually results in the $Fe_2As_2$ layer covered by atoms from the spacer, thus difficult to study the role a spacer plays using surface sensitive techniques. Recently, a new Fe-based superconducting system $Ca_{10}Pt_nAs_8(Fe_2As_2)_5$ with n = 3 (Ca10-3-8) and n = 4 (Ca10-4-8) has been reported[10, 11, 12]. As shown in Fig. 1a there is additional $Pt_nAs_8$ layer sandwiched in between two Ca layers within adjacent $Fe_2As_2$ building blocks: Ca10-3-8 is triclinic, while three different structures—tetragonal, triclinic, and monoclinic—have been reported for the Ca10-4-8 compound[10, 11, 12, 13]. According to first-principles calculations[11, 14, 15] and an angle-resolved photoemission spectroscopy study[16], these $Pt_nAs_8$ layers are conducting, in contrast to insulating spacers in 111, 1111, 122 families, and more complex compounds such as $Sr_3Sc_2O_5Fe_2As_2$[17] and $Sr_4V_2O_6Fe_2As_2$[18]. Thus, this new $Ca_{10}Pt_nAs_8(Fe_2As_2)_5$ system offers an excellent platform to explore the role of interlayer spacers in interlayer superconductivity.

In this paper, we report experimental investigations of superconducting $Ca_{10}Pt_4As_8(Fe_{2-x}Pt_xAs_2)_5$ single crystals layer-by-layer using scanning tunneling microscopy and spectroscopy (STM/S). The layer-by-layer probing of these layered superconductors is extremely important as illustrated in the recent study of $Bi_2Sr_2CaCu_2O_{8+\delta}$, where it was found that "the well-known pseudogap feature observed by STM is inherently property of the BiO planes and thus irrelevant directly to Cooper pairing"[19]. By creating a fresh surface through single-crystal cleavage under ultra-high vacuum, one can study both structural (STM topography) and electronic properties (scanning tunneling spectroscopy (STS)) of the exposed surfaces. By comparing STS spectra taken on different layers, we address how the surface structure and Ca concentration affect superconductivity on the surfaces of Ca10-4-8.

**Results**

For the $Ca_{10}Pt_4As_8(Fe_{2-x}Pt_xAs_2)_5$ single crystals we used for this study, an atomically-defined layered structure is clearly imaged by high-angle annular dark field scanning transmission electron microscopy (HAADF-STEM) taken along [210] direction (Fig. 1b). Similar to previous report[12], our crystals form a tetragonal structure with a lattice parameter $a$ = 8.733 Å, and interlayer distances indicated in Fig. 1a. The intensity of HAADF-STEM images strongly depends on the averaged atomic number (Z) in the projected atomic columns. As shown in Fig. 1b, the higher intensity of Fe (Z = 26) column than that of As (Z = 33) column indicates that there exists some Pt in $Fe_{2-x}Pt_xAs_2$ layer, i.e. $x \neq 0$. Noticeable Pt doping in $Fe_{2-x}Pt_xAs_2$ layers was reported in previous studies[10, 12], which is responsible for various reported $T_c$ values of the material. For our study, $Ca_{10}Pt_4As_8(Fe_{2-x}Pt_xAs_2)_5$ single crystals exhibit a superconducting transition at $T_c$ = 34 K, as shown in both the in-plane and out-of-plane resistivity in Fig. 1c.

Fresh surfaces were created by cleaving single crystals under ultrahigh vacuum (~ $10^{-10}$ torr) at ~ 90 – 100 K. Since Ca atoms are weakly bounded to adjacent layers but closer to the $Fe_{2-x}Pt_xAs_2$ layer than the $Pt_4As_8$ layer (see Fig. 1a), naively two surface terminations should be expected with roughly the same probability: (1) a full Ca layer on the top of $Fe_{2-x}Pt_xAs_2$ and (2) a bare $Pt_4As_8$ layer. Bulk-truncated $Pt_4As_8$ layer and $Fe_{2-x}Pt_xAs_2$ layer with their relative surface unit cells are shown in Figs. 1d – e, respectively. The $Fe_{2-x}Pt_xAs_2$ surface has a (1×1) structure with $a_{Fe_2As_2} = 3.9$ Å. Since the $x$ value is yet to be determined, showing Fig. 1e is the structure of $Fe_2As_2$ without indicating the location of Pt. (marked with a red "square" (□) represents a (1×1) structure in Fig. 1e). The $Pt_4As_8$ surface has a different (1×1) structure with $a_{Pt_4As_8} = 8.73$ Å (marked with a green "square" (□)), which has the same size as the primitive unit cell for bulk-truncation (Fig. 1a). This green "square" crystal unit cell for $Pt_4As_8$ is commensurate with the surface (1×1) $Fe_{2-x}Pt_xAs_2$ lattice. To facilitate consistent notation every structure will be denoted relative to the (1×1) $Fe_{2-x}Pt_xAs_2$ surface unit cell: the bulk-truncated (1×1) $Pt_4As_8$ structure (Fig. 1d) will be denoted as the Pt1-$\sqrt{5}\times\sqrt{5}$ structure. All observed structures are summarized in Table 1, which will be presented and discussed later.

Figure 2a shows the STM image of cleaved surfaces, displaying large flat terraces. Fig. 2b shows the line profile recorded along the marked location in Fig. 2a, where the step height is ~ 10 Å, which corresponds to a reported unit cell height[11, 12]. In addition to these large steps, there are regions that show much smaller steps (~ 1 Å). These are marked with arrows in Figs. 2a – b. The magnified image of such an area is shown in Fig. 2c. Notably, it consists of two terraces (A and B) with a height difference of only ~ 1 Å (see Fig. 2d). The observed surfaces can be identified using structural information displayed in Fig. 1a. Fig. 3a shows the topography of region A marked in Fig. 2c. Note it has a "square" unit cell with a lattice size ~ 8.7 Å with some local electronic inhomogeneity (Supplementary Fig. S1). This is identical to the lattice size of the bulk-truncated (1×1) $Pt_4As_8$ structure (Fig. 1d, "Pt1-$\sqrt{5}\times\sqrt{5}$"). Thus, we conclude that the region A is the $Pt_4As_8$ layer without reconstruction nor the presence of Ca. As shown in Figs. 1a and 1d, there are four Pt sites in the crystal unit cell: Pt1 is above the plane formed by Pt2 and Pt3, and Pt4 is below the plane. STM should observe primarily the Pt1 atoms, forming the "square" unit cell with the edge length of ~ 8.7 Å, as illustrated in Fig. 1d (green square) and labeled Pt1-$\sqrt{5}\times\sqrt{5}$.

Since the step height between regions A ($Pt_4As_8$) and B is ~ 1 Å, the region B can only be a Ca covered area on the top of $Pt_4As_8$. According to Fig. 1a, there are five Ca atoms on one Pt1-$\sqrt{5}\times\sqrt{5}$ unit cell: four (Ca1) are located at $z/c = 0.2418$ and one (Ca2) at $z/c = 0.2294$ (where $z/c = 0$ is the plane formed by Pt2 and Pt3)[12]. As illustrated in Fig. 3d, the weakly bounded Ca layer with a full-monolayer (ML) coverage should have a (1×1) structure with the lattice parameter of 3.9 Å (where 1 ML is defined as the amount of Ca atoms: four Ca1 and one Ca2 per primitive unit cell), the same structure as the bulk-truncated (1×1) $Fe_{2-x}Pt_xAs_2$ (Fig. 1e). However, the region B (see Fig. 3b) shows a different structure than the one with a full ML coverage. The Fourier transform pattern in the inset of Fig. 3b shows the same spots as observed in Pt1-$\sqrt{5}\times\sqrt{5}$ (see the inset of Fig. 3a): Ca2 atoms form a "square" unit cell with the lattice size of ~ 8.7 Å, which is commensurate with the Pt1-$\sqrt{5}\times\sqrt{5}$ underneath. While we expect to observe a fully Ca covered $Fe_{2-x}Pt_xAs_2$ layer and a bare $Pt_4As_8$ layer, the existence of Ca2-$\sqrt{5}\times\sqrt{5}$ can be traced back to the Ca distribution along the $c$ axis. Ca2 is ~ 0.15 Å closer to the $Pt_4As_8$ surface compared to Ca1, and is located on the top of Pt4 (Figs. 1a and 3d). When the crystal is cleaved, Ca2 atoms remain on the $Pt_4As_8$ surface while Ca1 atoms are peeled off. The resultant Ca2-$\sqrt{5}\times\sqrt{5}$ unit cell is marked in Fig. 3d with the red dashed square. By counting atoms shown in Fig. 3b, the Ca coverage is estimated to be ~ 0.17 ML, which is close to 0.2 ML, the amount of Ca2 to form a perfect $\sqrt{5}\times\sqrt{5}$ structure (Ca1:Ca2 = 4:1). However, the additional weak and broad spots marked with arrows in the Fourier transform patterns (see the inset of Fig. 3b) indicate that there is additional short-range order formed by Ca atoms.

In addition to the bare Pt1-$\sqrt{5}\times\sqrt{5}$ (region A) and Ca2-$\sqrt{5}\times\sqrt{5}$ (regions B) surfaces, we observe another surface. As shown in Fig. 3c, this surface does not form any ordered structure. As will be discussed later, this is identified as a Ca covered $Fe_{2-x}Pt_xAs_2$ layer ("Ca-disordered").

Returning to the Ca2-$\sqrt{5}\times\sqrt{5}$ surface, we now illustrate that it is metastable. In the scanned area of ~ 9 $\mu m^2$ from 3 different as-cleaved samples, Pt1-$\sqrt{5}\times\sqrt{5}$ surface (86 %) and Ca2-$\sqrt{5}\times\sqrt{5}$ layer (14 %) on top of Pt1-$\sqrt{5}\times\sqrt{5}$ surface are observed. In some Ca2-$\sqrt{5}\times\sqrt{5}$ regions, Fourier transform patterns show stronger additional spots (marked with arrows in the inset of Fig. 3b) with weaker $\sqrt{5}\times\sqrt{5}$ spots, indicating the Ca2-$\sqrt{5}\times\sqrt{5}$ surface is not a ground state. Fig. 4a shows the Ca layer after the sample is warmed from 4.3 K to room temperature (~ 290 K) for an hour, revealing a new structure with a $\sqrt{10}\times\sqrt{10}$ unit cell ("Ca2-$\sqrt{10}\times\sqrt{10}$") displayed in Figs. 4a – b. Ca2-$\sqrt{10}\times\sqrt{10}$ is the ($\sqrt{2}\times\sqrt{2}$)R45° reconstruction of the original Ca2-$\sqrt{5}\times\sqrt{5}$ unit cell, suggesting the Ca surface undergoes a structure change during the warming process. Since the Ca coverage for the perfect Ca2-$\sqrt{5}\times\sqrt{5}$ structure is 0.2 ML, there are excess Ca atoms (0.1 ML) after forming Ca2-$\sqrt{10}\times\sqrt{10}$ structure. These excessive Ca atoms form clusters on the surface as shown in Fig. 4a. With an additional hour-long annealing at room temperature, all surface Ca atoms are clustered ("Ca-clustered"), revealing the Pt1-$\sqrt{5}\times\sqrt{5}$ surface underneath (Fig. 4c and the region C in Fig. 4d): the square lattice of Pt1-$\sqrt{5}\times\sqrt{5}$ surface can be seen in between clusters (one of such area is marked with a yellow circle in Fig. 4c). Thus, the Ca-clustered surface is stable and the Ca2-$\sqrt{5}\times\sqrt{5}$ and Ca2-$\sqrt{10}\times\sqrt{10}$ surfaces are metastable.

As for the Ca-disordered $Fe_{2-x}Pt_xAs_2$ surface (Fig. 3c), it remains disordered after annealing (Fig. 4d). According to the line profile shown in Fig. 4e, the step height between regions C and D is ~ 3 Å. Since the region C is the $Pt_4As_8$ layer, the region D must be the Ca layer on the top of $Fe_{2-x}Pt_xAs_2$ layer. We also observed the same disordered surface located about 3.5 Å lower than the Ca2-$\sqrt{10}\times\sqrt{10}$ surface (Supplementary Fig. S2), after initial warming up from 4.3 K to 290 K. The morphological feature is also the same as that shown in Fig. 3c, which is obtained from the as-cleaved Ca/$Fe_{2-x}Pt_xAs_2$ surface. For comparison, the surface of $CaFe_2As_2$ has a (1×2) structure formed by 0.5 ML Ca on the top of $Fe_2As_2$ layer[20]. For $Ca_{10}Pt_4As_8(Fe_{2-x}Pt_xAs_2)_5$, the absence of ordered Ca/$Fe_{2-x}Pt_xAs_2$ surface is undoubtedly due to the asymmetric interlayer spacing: the Ca layer is closer to the $Fe_{2-x}Pt_xAs_2$ layer than to the $Pt_4As_8$ layer. When the sample is cleaved, most of Ca atoms stay on the top of $Fe_{2-x}Pt_xAs_2$ layer, which are disordered as shown in Figs. 3c (as-cleaved) and 4d (after annealing). In contrast to the Ca/$Pt_4As_8$ surface, the disordered Ca/$Fe_{2-x}Pt_xAs_2$ surface is insensitive to annealing. Only the surface corrugation changes from ~ 1 Å (as-cleaved) to ~ 2 Å (after annealing). The large amount of disordered Ca atoms prevents from seeing the bare $Fe_{2-x}Pt_xAs_2$ layer.

Cleaving a $Ca_{10}Pt_4As_8(Fe_{2-x}Pt_xAs_2)_5$ single crystal results in three surfaces and two additional surfaces via annealing as summarized in Table I. In the Ca2-$\sqrt{5}\times\sqrt{5}$ surface, the step height (~ 1 Å) observed in Fig. 2d is ~ 0.7 Å shorter than the corresponding bulk spacing between Ca2 and Pt1 (1.67 Å). This STM result is consistent with the behavior reported at the surface of $CaFe_2As_2$ with 0.5 ML Ca coverage[20], where Ca atoms are pulled down by ~ 0.5 Å (~ 30 %) determined by low energy electron diffraction (LEED). In addition, the calculated charge states using the Bader scheme in Ref. 15 [$(Ca_{10}^{14.144+})$, $(Fe_2As_2)_5^{1.899-}$, and $(Pt_4As_8)^{12.252-}$] can be utilized to support the large inward relaxation. In bulk, positively charged Ca atoms are positioned in between two negatively charged layers, $Fe_{2-x}Pt_xAs_2$ and $Pt_4As_8$. When the negatively charged $Fe_{2-x}Pt_xAs_2$ layer is removed by cleaving there would be too many positive charged Ca atoms unless some are removed with the $Fe_{2-x}Pt_xAs_2$ layer. If the remaining Ca is indeed above Pt4 atom, this specific Ca atom (Ca2) has more room to be pulled down than other four Ca atoms (Ca1) since Pt4 is located at 0.72 Å below the plane formed by Pt2 and Pt3 at $z/c = 0$. In the Ca-disordered surface, the remaining Ca atoms do not form any ordered structure as shown in Figs. 3c and 4d. This may be the reason that the observed spacing (~ 3 Å) between $Pt_4As_8$ layer and Ca covered $Fe_{2-x}Pt_xAs_2$ layer (Fig. 4e) is close to the spacing in the bulk (3.25 Å).

The most significant observation on the cleaved Ca10-4-8 surfaces comes from the STS measurements. From the bulk measurements, $T_c$ of Ca10-4-8 is ~ 34 K (see Fig. 1c). Thus, STS measurements at a much lower temperature (4.3 K) than $T_c$ are expected to show the opening of superconducting energy gap. This is indeed observed on the ordered Ca surfaces on $Pt_4As_8$. Figs. 5a – b show spectra taken on the Ca2-$\sqrt{5}\times\sqrt{5}$ and Ca2-$\sqrt{10}\times\sqrt{10}$ surfaces with corresponding STM images (insets). The annealed surface (Ca2-$\sqrt{10}\times\sqrt{10}$) shows enhanced superconducting coherence peaks (Fig. 5b) when compared to the spectrum taken on the as-cleaved surface (Ca2-$\sqrt{5}\times\sqrt{5}$) (Fig. 5a). Note that the spectra shown in Fig. 5 are raw data (no normalization) so information about absolute conductance cannot be obtained. However, from the coherence peak positions in Fig. 5b, one can estimate the energy gap $\Delta$ ~ 4.4 meV at T = 4.3 K. This results in that $2\Delta(T = 4.3 \text{ K})/k_BT_c = 3.0$. According to the BCS theory, $2\Delta(T = 0 \text{ K})/k_BT_c = 3.5$ for weakly coupled superconductors. The smaller $2\Delta/k_BT_c$ may result from underestimated $\Delta$ due to finite temperature and/or overestimated $T_c$ at the surface. It is worth pointing out that the spectrum (Fig. 5b) exhibits a finite zero bias conductance (ZBC). Such a feature has been observed previously in other Fe-based superconductors[20, 21, 22, 23, 24, 25, 26], even when probed at much lower temperatures than their $T_c$, indicating the ZBC is not due to the thermal broadening effect. It was also shown that spectral features can vary on different reconstructed surfaces in $BaFe_{2-x}Co_xAs_2$[26] suggesting reconstructions may be closely related to the observed ZBC in Fe-base superconductors. The origin of the ZBC, however, is not clear at the moment and requires further investigation.

The STS data for the Pt1-$\sqrt{5}\times\sqrt{5}$ surface shows a kink with positive bias but smooth $dI/dV$ with negative bias (Fig. 5c). In addition, the detailed feature in each spectrum varies significantly depending on the site where STS is taken (Fig. S4). Likewise, the STS spectra taken on the Ca-disordered (on the $Fe_{2-x}Pt_xAs_2$) surface lack any features that could be associated with superconductivity (Fig. 5d). These spectra are strongly site dependent as well.

**Discussion**

What preserves superconductivity on the Ca-ordered $Pt_4As_8$ surfaces (Ca2-$\sqrt{5}\times\sqrt{5}$ and Ca2-$\sqrt{10}\times\sqrt{10}$) but diminishes superconductivity on the bare $Pt_4As_8$ surface (Pt1-$\sqrt{5}\times\sqrt{5}$) and Ca-disordered $Fe_{2-x}Pt_xAs_2$ surface? It is well known that the $Fe_2As_2$ layer is responsible for superconductivity in bulk Fe-based superconductors, and the $Pt_4As_8$ layer is expected to be superconducting as well, given the nature of 3D superconductivity (Fig. 1c). We recall that superconducting coherence peaks have been observed in spectra taken on the surfaces of other Fe-based superconductors[20, 23, 27], in which the surfaces are ordered. For example, the cleaved surface of $Ca(Fe_{0.925}Co_{0.075})_2As_2$ has an ordered stripe phase formed by a half monolayer Ca on the $Fe_2As_2$ layer[20]. Moreover, superconducting features are enhanced in the Ca2-$\sqrt{10}\times\sqrt{10}$ (Fig.

5b) with a better ordered surface than the Ca2-$\sqrt{5}\times\sqrt{5}$, as seen in the Fourier transform pattern of the Ca2-$\sqrt{10}\times\sqrt{10}$, which is much sharper than the one of Ca2-$\sqrt{5}\times\sqrt{5}$ (see insets of Fig. 3b and 4a). This suggests that an ordered surface favors superconductivity in this system. Thus, the absence of superconducting coherence peak in our STS spectra obtained from Ca-disordered $Fe_{2-x}Pt_xAs_2$ is due to the disorder of Ca atoms. However, this cannot explain why there is no coherence peaks seen in the bare $Pt_4As_8$ layer, where we observed an ordered structure (Fig. 5c). Here we propose that the lack of superconductivity on a bare $Pt_4As_8$ layer is a result of charge imbalance. In comparison with Ca2-$\sqrt{5}\times\sqrt{5}$ and Ca2-$\sqrt{10}\times\sqrt{10}$ surfaces, the bare $Pt_4As_8$ layer has no Ca atoms, which must be responsible for the absence of superconducting coherence peaks. In the bulk, the charge transfer from Ca layer to $Pt_4As_8$ layer is much greater than the charge transfer from Ca to $Fe_{2-x}Pt_xAs_2$ layer[15]. The lack of Ca makes the Pt1-$\sqrt{5}\times\sqrt{5}$ surface not superconducting which is much different than the bulk. Compared to the bare Pt1-$\sqrt{5}\times\sqrt{5}$ surface, the Ca2-$\sqrt{5}\times\sqrt{5}$ and Ca2-$\sqrt{10}\times\sqrt{10}$ surfaces have much less severe charge imbalance, thus preserving superconductivity. The effect of charge imbalance is also evidenced at the surface of $Sr_{1-x}K_xFe_2As_2$[27], where no sign of superconducting coherence peak is observed on the ordered bare $Fe_2As_2$ layer. Moreover, it was shown that superconductivity can be tuned by charge transfer from gated ionic liquid to FeSe[28].

In view of the previous studies of Fe-based superconductors including 11, 111, 122 systems, cleaving creates two identical surfaces, due to structure symmetry. A material like $Ca_{10}Pt_4As_8(Fe_{2-x}Pt_xAs_2)_5$ is fundamentally different offering great opportunities to study the structural and physical properties layer by layer, and provide important implication on the fabrication of new Fe-based superconductors. We have demonstrated that the electronic properties of the surface depend dramatically upon the stoichiometry and the arrangement of surface atoms. Cleaving and processing of the single crystal sample produce five distinct surface phase: a bare $Pt_4As_8$ surface (Pt1-$\sqrt{5}\times\sqrt{5}$), an ordered 0.2 ML Ca array on $Pt_4As_8$ surface (Ca2-$\sqrt{5}\times\sqrt{5}$), an ordered 0.1 ML Ca on $Pt_4As_8$ (Ca2-$\sqrt{10}\times\sqrt{10}$), a Ca-clustered $Pt_4As_8$ layer, and a Ca-disordered layer on $Fe_{2-x}Pt_xAs_2$.

The fact that STS reveals no sign for superconductivity on the bare $Pt_4As_8$ surface and the disordered Ca on $Fe_{2-x}Pt_xAs_2$ indicates that both structure order and charge balance are crucial for superconductivity. We believe the unequal distance of Ca atoms with respect to both $Fe_{2-x}Pt_xAs_2$ layer and $Pt_4As_8$ layer is the reason for the existence of these strange surface structures. When the surface contains ingredients sufficient to maintain charge balance, properties seen in bulk can be preserved on the surface. This is supported by our observation of superconducting coherent peaks in STS on the Ca2-$\sqrt{5}\times\sqrt{5}$ and Ca2-$\sqrt{10}\times\sqrt{10}$ surfaces of $Ca_{10}Pt_4As_8(Fe_{2-x}Pt_xAs_2)_5$.

While three-dimensional superconductivity in layered superconductors is considered through tunneling between layers, our STM/STS study of $Ca_{10}Pt_4As_8(Fe_{2-x}Pt_xAs_2)_5$ reveals that proximity effect would not maintain the superconductivity on the surface if the surface experiences different chemical environment. As can be seen in Fig. 4e, both bare $Pt_4As_8$ layer and the disordered Ca layer are just about 3 – 4 Å from the $Fe_{2-x}Pt_xAs_2$ layer. The absence of superconductivity in these layers indicates that the top most $Fe_{2-x}Pt_xAs_2$ layer is not superconducting, even though the adjacent $Fe_{2-x}Pt_xAs_2$ layer is only ~ 8 Å below (Fig. 1a).

In summary, we have investigated the detailed surface structures and electronic properties of $Ca_{10}Pt_4As_8(Fe_{2-x}Pt_xAs_2)_5$ using STM/STS. We observe five different surfaces as summarized in Table 1. Remarkably, STS measurements reveal that only spectra taken on the Ca surfaces with *ordered structures* and *appropriate amount* show superconducting coherence peaks with finite zero bias conductance. Neither bare $Pt_4As_8$ surface nor disordered $Ca/Fe_{2-x}Pt_xAs_2$ exhibit superconductivity. Our results indicate that superconductivity can only be preserved on the surface when it has similar chemical arrangement as the bulk counterpart: either missing or disordered Ca would kill superconductivity in the surface planes, due to charge imbalance. Furthermore, our results confirm that the intermediate $Pt_4As_8$ layer is superconducting in $Ca_{10}Pt_4As_8(Fe_{2-x}Pt_xAs_2)_5$, direct evidence for bulk superconductivity.

**Methods**

**Single crystal growth.** High quality single crystals of Ca10-4-8 were grown using the self-flux method. The stoichiometric amounts of high purity calcium shot (99.999 % Alfa Aesar), Pt powder (99.95 % Alfa Aesar), iron powder (99.95 % Alfa Aesar), and arsenic powder (99.999 % Alfa Aesar) are mixed in the ratio 10:4:10:18. The mixture is placed in an alumina crucible and sealed in a quartz tube under vacuum. The whole assembly is heated in a box furnace to 700 °C at a rate of 150 °C/h and is held at this temperature for 5 h. It is further heated to 1100 °C at a rate of 80 °C/h where it is held for 50 h, and then cooled to 1050 °C at a rate of 1.25 °C/h. It is further cooled to 500 °C at a rate of 5.5 °C/h and finally cooled down to room temperature by turning off the power. Shiny black plate-like single crystals are obtained without requiring any additional process. These crystals have a typical size of $6 \times 6 \times 0.2$ mm$^3$.

**STM/S measurements.** Ca10-4-8 single crystals are first cleaved at low temperatures (~ 90 – 100 K) then transferred to the scanning stage either at 80 K or 4.3 K. $10^{10}$ V/A (current amplifier gain) is used at 4.3 K and $10^9$ V/A at 80 K so that different tunneling set currents are used at 4.3 K (2 – 40 pA) and 80 K (100 pA). No noticeable difference is observed in the surface topology taken at 80 K and 4.3 K. $dI/dV$ spectra are acquired by using a lock-in amplifier with $V_{mod}$ =

80µV and $f_{mod}$ = 555 Hz at 4.3 K. WSxM software was among the tools used for image preparation and analysis[29].

**STEM measurements.** STEM imaging is performed using a JEM-ARM200F microscope equipped with probe-corrector operating at 200 kV. Samples for TEM observation along different axis-zone direction are prepared by using Focused Ion Beam (FIB).

**Acknowledgements**

This work was supported by NSF-DMR 1504226 (J. K., G. L., A. K., R. J., and E. W. P.), ONR-N00014-14-1-0330 (H. N. and C. K. S.), DOE DE-SC0002136 (Z. W. and J. Z.). The STEM measurement was conducted in Brookhaven National Laboratory which was supported by the U.S. Department of Energy, Office of Basic Energy Science, under Contract No. DE-AC02-98CH10886 (Y. Z.).

**Author contributions**

R. J. and E. W. P. proposed and designed the research. A. B. K. and R. J. contributed to single crystal growth. J. K., H. N. and G. L. carried out the STM/STS experiment with the assistance from C. K. S.. Z. W. and Y. Z. carried out the STEM experiment. J. K, J. Z., R. J. and E. W. P. analyzed the data. J. K. wrote the paper with J. Z., R. J. and E. W. P.

**Competing financial interests**: The authors declare no competing financial interests.

Table 1: Summary of observed surfaces and their structures. The surfaces where superconductivity is observed are also marked.

|  | Observed Surface | Referred Name | Lattice Size | SC |
|---|---|---|---|---|
| as-cleaved | bare $Pt_4As_8$ surface (no reconstruction): $\sqrt{5}\times\sqrt{5}$ structure | "Pt1-$\sqrt{5}\times\sqrt{5}$" | 8.7 Å: a square unit cell made of four Pt1 atoms | No |
|  | reconstructed $\sqrt{5}\times\sqrt{5}$ Ca surface on the $Pt_4As_8$ layer | "Ca2-$\sqrt{5}\times\sqrt{5}$" | 8.7 Å: a square unit cell made of four Ca2 atoms | Yes |
|  | disordered Ca surface on the $Fe_2As_2$ layer | "Ca-disordered" |  | No |
| after warming up to RT | Reconstructed $\sqrt{10}\times\sqrt{10}$ Ca surface on the $Pt_4As_8$ layer | "Ca2-$\sqrt{10}\times\sqrt{10}$" | 12.3 Å: $(\sqrt{2}\times\sqrt{2})$ R45° reconstruction of "Ca2-$\sqrt{5}\times\sqrt{5}$" | Yes |
| after additional hour long annealing at RT | Ca clusters on the $Pt_4As_8$ layer | "Ca-clustered" |  |  |

# References

1. Tapp, J. H. *et al.* LiFeAs: An intrinsic FeAs-based superconductor with $T_c$=18 K. *Phys. Rev. B* **78,** 060505 (2008).

2. Kamihara, Y., Watanabe, T., Hirano, M. & Hosono, H. Iron-based layered superconductor La[$O_{1-x}F_x$]FeAs (x = 0.05−0.12) with $T_c$ = 26 K. *J. Am. Chem. Soc.* **130,** 3296-3297 (2008).

3. Matsuishi, S. *et al*. Superconductivity induced by Co-doping in quaternary fluoroarsenide CaFeAsF. *J. Am. Chem. Soc.* **130,** 14428-14429 (2008).

4. Rotter, M., Tegel, M. & Johrendt, D. Superconductivity at 38 K in the iron arsenide ($Ba_{1-x}K_x$)$Fe_2As_2$. *Phys. Rev. Lett.* **101,** 107006 (2008).

5. Altarawneh, M. M. *et al*. Determination of anisotropic $H_{c2}$ up to 60 T in $Ba_{0.55}K_{0.45}Fe_2As_2$ single crystals. *Phys. Rev. B* **78,** 220505 (2008).

6. Yamamoto, A. *et al.* Small anisotropy, weak thermal fluctuations, and high field superconductivity in Co-doped iron pnictide Ba($Fe_{1-x}Co_x$)$_2As_2$. *Appl. Phys. Lett.* **94,** 062511 (2009).

7. Yuan, H. Q. *et al.* Nearly isotropic superconductivity in (Ba,K)$Fe_2As_2$. *Nature* **457,** 565-568 (2009).

8. Cho, K. *et al.* Anisotropic upper critical field and possible Fulde-Ferrel-Larkin-Ovchinnikov state in the stoichiometric pnictide superconductor LiFeAs. *Phys. Rev. B* **83,** 060502 (2011).

9. Jo, Y. J. *et al.* High-field phase-diagram of Fe arsenide superconductors. *Physica C* **469,** 566-574 (2009).

10. Kakiya, S. *et al.* Superconductivity at 38 K in iron-based compound with platinum-arsenide layers $Ca_{10}(Pt_4As_8)(Fe_{2-x}Pt_xAs_2)_5$. *J. Phys. Soc. Jpn.* **80,** 093704 (2011).

11. Löhnert, C. *et al*. Superconductivity up to 35 K in the iron platinum arsenides (CaFe$_{1−x}$Pt$_x$As)$_{10}$Pt$_{4−y}$As$_8$ with layered structures. *Angew. Chem. Int. Ed.* **50,** 9195-9199 (2011).

12. Ni, N., Allred, J. M., Chan, B. C. & Cava R. J. High $T_c$ electron doped $Ca_{10}(Pt_3As_8)(Fe_2As_2)_5$ and $Ca_{10}(Pt_4As_8)(Fe_2As_2)_5$ superconductors with skutterudite intermediary layers. *Proc. Natl. Acad. Sci. U.S.A.* **108,** E1019-E1026 (2011).

13. Stürzer, T., Derondeau, G. & Johrendt, D. Role of different negatively charged layers in $Ca_{10}(FeAs)_{10}(Pt_4As_8)$ and superconductivity at 30 K in electron-doped ($Ca_{0.8}La_{0.2}$)$_{10}(FeAs)_{10}(Pt_3As_8)$. *Phys. Rev. B* **86,** 060516 (2012).

14. Nakamura, H. & Machida, M. First-principles study of Ca–Fe–Pt–As-type iron-based superconductors. *Physica C* **484,** 39-42 (2013).

15. Shein, I. R. & Ivanovskii, A. L. Ab initio study of the nature of the chemical bond and electronic structure of the layered phase $Ca_{10}(Pt_4As_8)(Fe_2As_2)_5$ as a parent system in the search for new superconducting iron-containing materials. *Theor. Exp. Chem.* **47,** 292-295 (2011).

16. Shen, X. P. *et al.* Electronic structure of $Ca_{10}(Pt_4As_8)(Fe_{2-x}Pt_xAs_2)_5$ with metallic $Pt_4As_8$ layers: An angle-resolved photoemission spectroscopy study. *Phys. Rev. B* **88,** 115124 (2013).

17. Zhu, X. *et al.* $Sr_3Sc_2Fe_2As_2O_5$ as a possible parent compound for FeAs-based superconductors. *Phys. Rev. B* **79,** 024516 (2009).

18. Zhu, X. *et al.* Transition of stoichiometric $Sr_2VO_3FeAs$ to a superconducting state at 37.2 K. *Phys. Rev. B* **79,** 220512 (2009).

19. Lv, Y. -F. *et al.* Mapping the electronic structure of each ingredient oxide layer of high-$T_c$ cuprate superconductor $Bi_2Sr_2CaCu_2O_{8+\delta}$. *Phys. Rev. Lett.* **115,** 237002 (2015).

20. Li, G. *et al.* Role of antiferromagnetic ordering in the (1 × 2) surface reconstruction of $Ca(Fe_{1-x}Co_x)_2As_2$. *Phys. Rev. Lett.* **112,** 077205 (2014).

21. Teague, M. L. *et al.* Measurement of a sign-changing two-gap superconducting phase in electron-doped $Ba(Fe_{1-x}Co_x)_2As_2$ single crystals using scanning tunneling spectroscopy. *Phys. Rev. Lett.* **106,** 087004 (2011).

22. Millo, O. *et al.* Scanning tunneling spectroscopy of $SmFeAsO_{0.85}$: Possible evidence for d-wave order-parameter symmetry. *Phys. Rev. B* **78,** 092505 (2008).

23. Yin, Y. *et al.* Scanning tunneling spectroscopy and vortex imaging in the iron pnictide superconductor $BaFe_{1.8}Co_{0.2}As_2$. *Phys. Rev. Lett.* **102,** 097002 (2009).

24. Massee, F. *et al.* Nanoscale superconducting-gap variations and lack of phase separation in optimally doped $BaFe_{1.86}Co_{0.14}As_2$. *Phys. Rev. B* **79,** 220517 (2009).

25. Massee, F. *et al.* Pseudogap-less high-$T_c$ superconductivity in $BaCo_xFe_{2-x}As_2$. *EPL (Europhysics Letters)* **92,** 57012 (2010).

26. Zhang, H. *et al.* Sqrt[2]×sqrt[2] structure and charge inhomogeneity at the surface of superconducting $BaFe_{2-x}Co_xAs_2$ (x=0–0.32). *Phys. Rev. B* **81,** 104520 (2010).

27. Song, C. -L. *et al.* Dopant clustering, electronic inhomogeneity, and vortex pinning in iron-based superconductors. *Phys. Rev. B* **87,** 214519 (2013).

28. Lei, B. *et al.* Evolution of high-temperature superconductivity from a low-$T_c$ phase tuned by carrier concentration in FeSe thin flakes. *Phys. Rev. Lett.* **116,** 077002 (2016).

29. Horcas, I. *et al.* WSXM: A software for scanning probe microscopy and a tool for nanotechnology. *Rev. Sci. Instrum.* **78,** 013705 (2007).

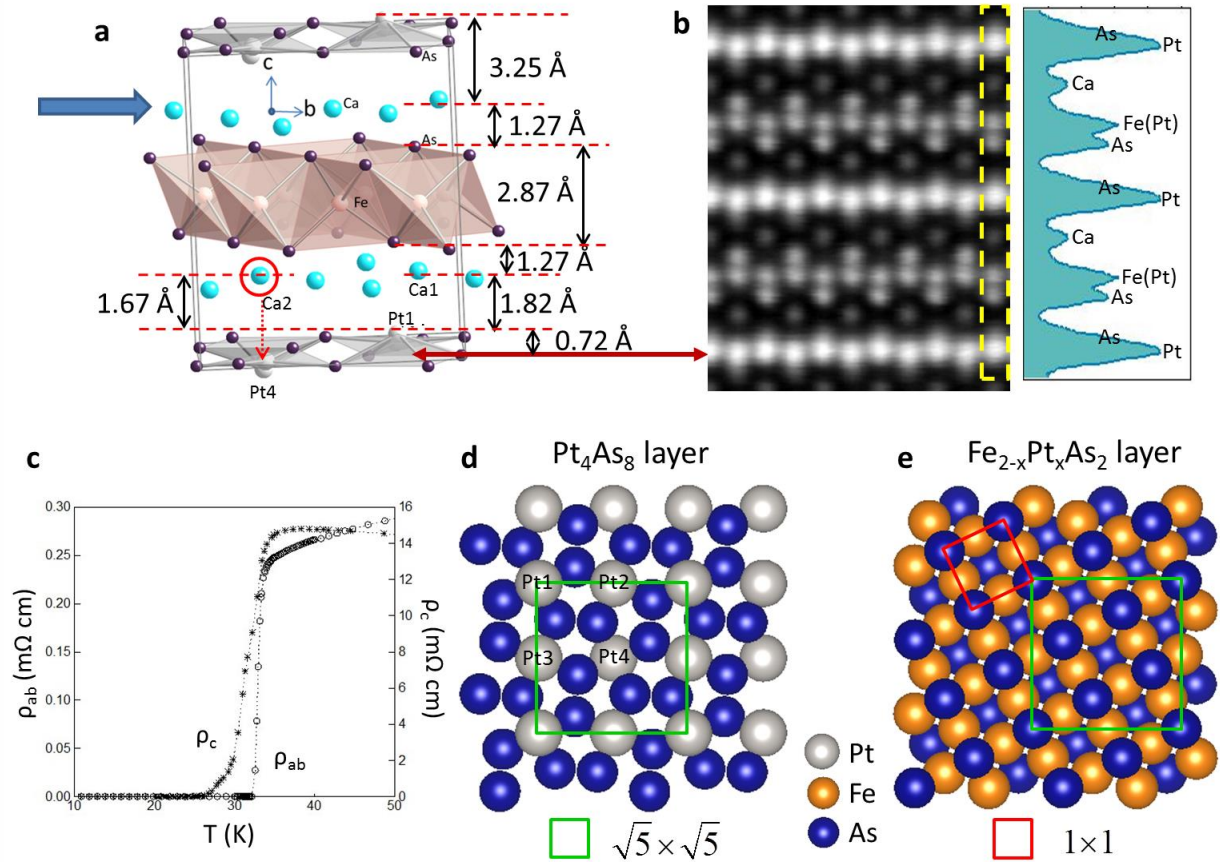

**Figure 1| Structure of the Ca10-4-8** (**a**) Schematic of the Ca10-4-8 bulk structure with the relative spacings between the different planes. Preferred cleavage plane is marked with a blue arrow. (**b**) HAADF-STEM image taken along [210] direction shows detailed structure of $Ca_{10}Pt_4As_8(Fe_{2-x}Pt_xAs_2)_5$: one of $Pt_4As_8$ layers and the corresponding plane in (a) are marked with a red arrow. Intensity profile measured from the yellow rectangle provides the configuration of atomic planes stacked along c direction: alternating Pt doped $Fe_2As_2$ and $Pt_4As_8$ layers with

Ca ions in between them. Double stack of the structure in Fig. 1a is shown in the image. (**c**) In-plane resistivity ($\rho_{ab}$) shows a clear superconducting transition with onset $T_c$ = 34 K and zero resistivity $T_c$ = 31 K. Out-of-plane resistivity ($\rho_c$) has slightly broad superconducting transition with a peak near 38 K. Bulk-truncated $Pt_4As_8$ layer (**d**) and $Fe_{2-x}Pt_xAs_2$ layer (**e**). The unit cell of each surface is marked with green (Pt1-$\sqrt{5}\times\sqrt{5}$) and red (1×1) squares.

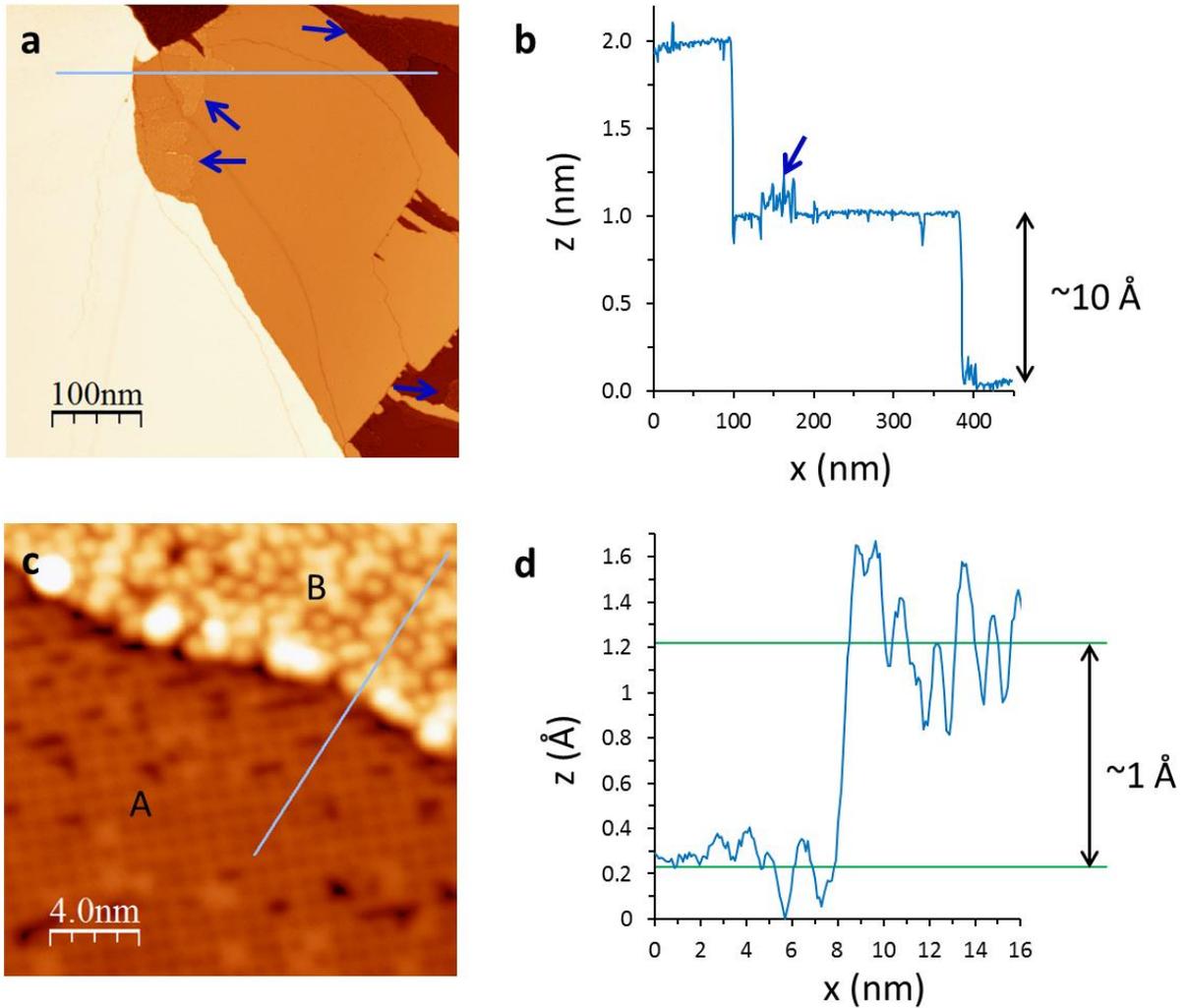

**Figure 2| Pt1-$\sqrt{5}\times\sqrt{5}$ surface and Ca2-$\sqrt{5}\times\sqrt{5}$ surface.** (**a** – **b**) STM image (taken at 4.3 K) of the stepped $Pt_4As_8$ surfaces ($V_{sample}$ = 0.7 V, I = 2.3 pA) and a line profile showing steps with a

reported unit cell height[11, 12]. Ca covered areas are marked with arrows. One of such Ca covered area is shown in (**c**): STM image (taken at 4.3 K) showing both Pt1-$\sqrt{5}\times\sqrt{5}$ surface (the region A) and Ca2-$\sqrt{5}\times\sqrt{5}$ surface (the region B) ($V_{sample}$ = 0.2 V, I = 2 pA). (**d**) Line profile shows Pt1-$\sqrt{5}\times\sqrt{5}$ layer with a small step height (~ 1 Å) to the Ca2-$\sqrt{5}\times\sqrt{5}$ surface.

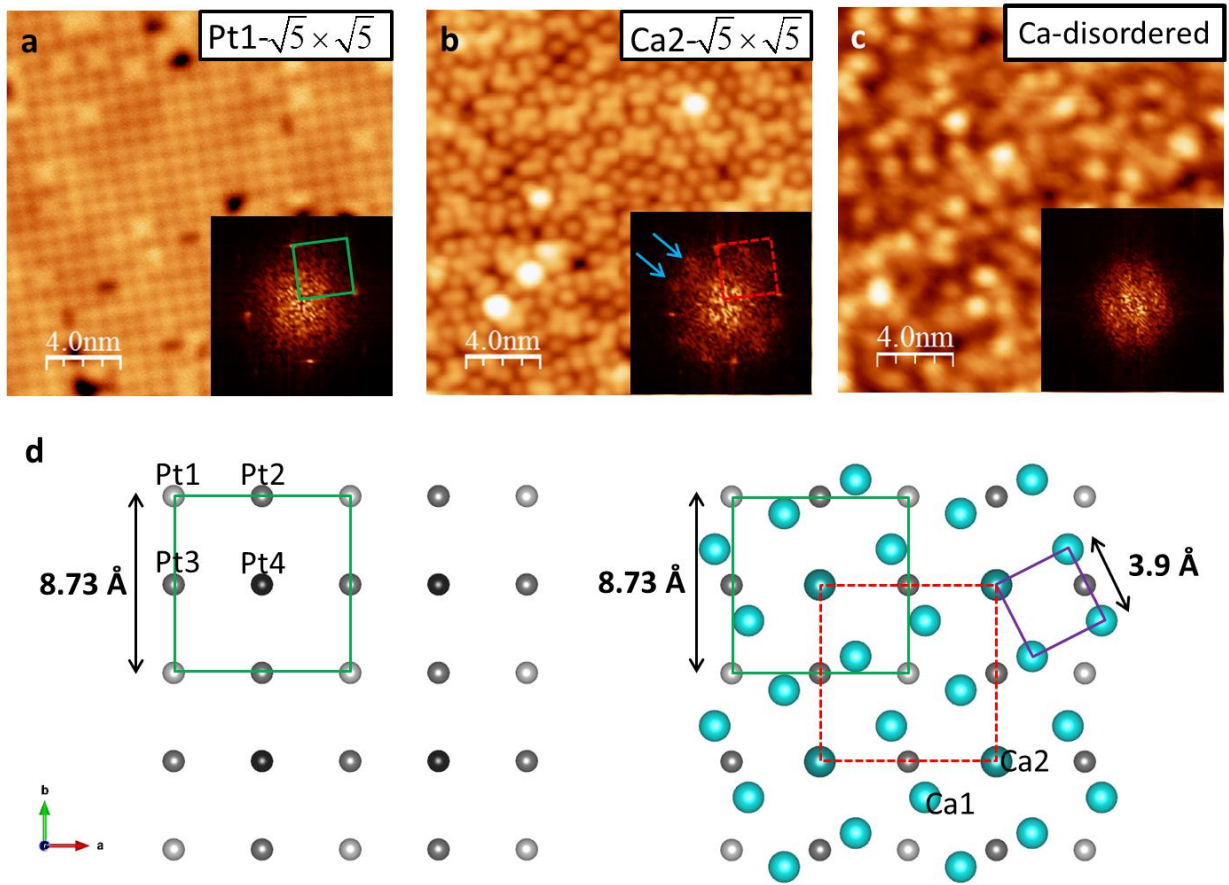

**Figure 3| Observed surfaces and their relative structures.** STM image of (**a**) Pt1-$\sqrt{5}\times\sqrt{5}$ surface ($V_{sample}$ = 0.7 V, I = 2.4 pA), (**b**) Ca2-$\sqrt{5}\times\sqrt{5}$ surface ($V_{sample}$ = 0.2 V, I = 2 pA), and (**c**) disordered Ca covered $Fe_{2-x}Pt_xAs_2$ layer ($V_{sample}$ = 0.2 V, I = 2.4 pA). All images are taken at 4.3 K. (insets) corresponding Fourier transform patterns. (**d**) Schematic shows relative relations of

each surface structure. The green solid square represents $\sqrt{5}\times\sqrt{5}$ unit cell of the Pt1-$\sqrt{5}\times\sqrt{5}$ surface (as also shown in Fig. 1d), the purple solid square represents (1×1) structure of Ca layer with a full-monolayer (ML) coverage, and the red dashed square is $\sqrt{5}\times\sqrt{5}$ unit cell of the Ca2-$\sqrt{5}\times\sqrt{5}$ surface. Pt1 (Pt4) is located at $z/c=0.06845$ ($-0.06845$) above (below) the Pt in-plane (Pt2, Pt3) at $z/c=0$. Ca2 atoms forming $\sqrt{5}\times\sqrt{5}$ unit cell are located at $z/c=0.2294$, which is ~ 0.15 Å lower than other Ca atoms (Ca1 at $z/c=0.2418$).

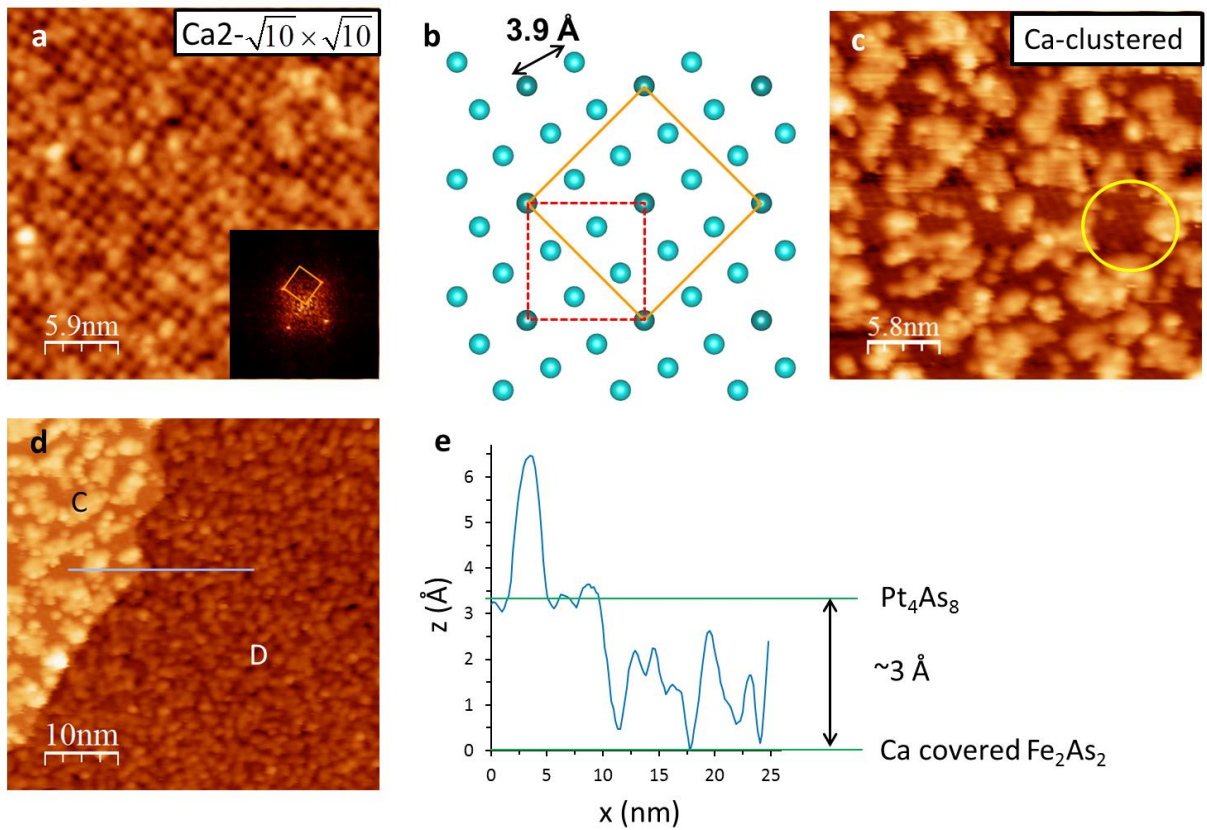

**Figure 4| Annealing effect on Ca2-$\sqrt{5}\times\sqrt{5}$ and disordered Ca surfaces**. (**a** – **b**) STM image (taken at 4.3 K) of Ca2-$\sqrt{10}\times\sqrt{10}$ surface after warmed from 4.3 K to room temperature (~ 290 K) and schematic of the structure ($V_{sample}$ = 100 mV, I = 2 pA). (inset) Fourier transform pattern. STM image of Ca2-$\sqrt{10}\times\sqrt{10}$ surface taken at 80 K is shown in Supplementary Fig. S3: there is

no significant difference between images taken at 4.3 K and 80 K. (**c**) Additional hour long annealing at room temperature causes surface Ca atoms become clustered, revealing underneath Pt1-$\sqrt{5}\times\sqrt{5}$ surface (taken at 80 K, $V_{sample}$ = 1 V, and I = 100 pA). One of such revealed Pt1-$\sqrt{5}\times\sqrt{5}$ surfaces is marked with a yellow circle. (**d**) STM image showing two different layers (taken at 80 K, $V_{sample}$ = 1 V, and I = 100 pA): Pt1-$\sqrt{5}\times\sqrt{5}$ with Ca clusters (the region C) and disordered Ca surface on $Fe_{2-x}Pt_xAs_2$ layer (the region D). (**e**) Distance between $Pt_4As_8$ layer and Ca covered $Fe_{2-x}Pt_xAs_2$ layer is shown in the line profile.

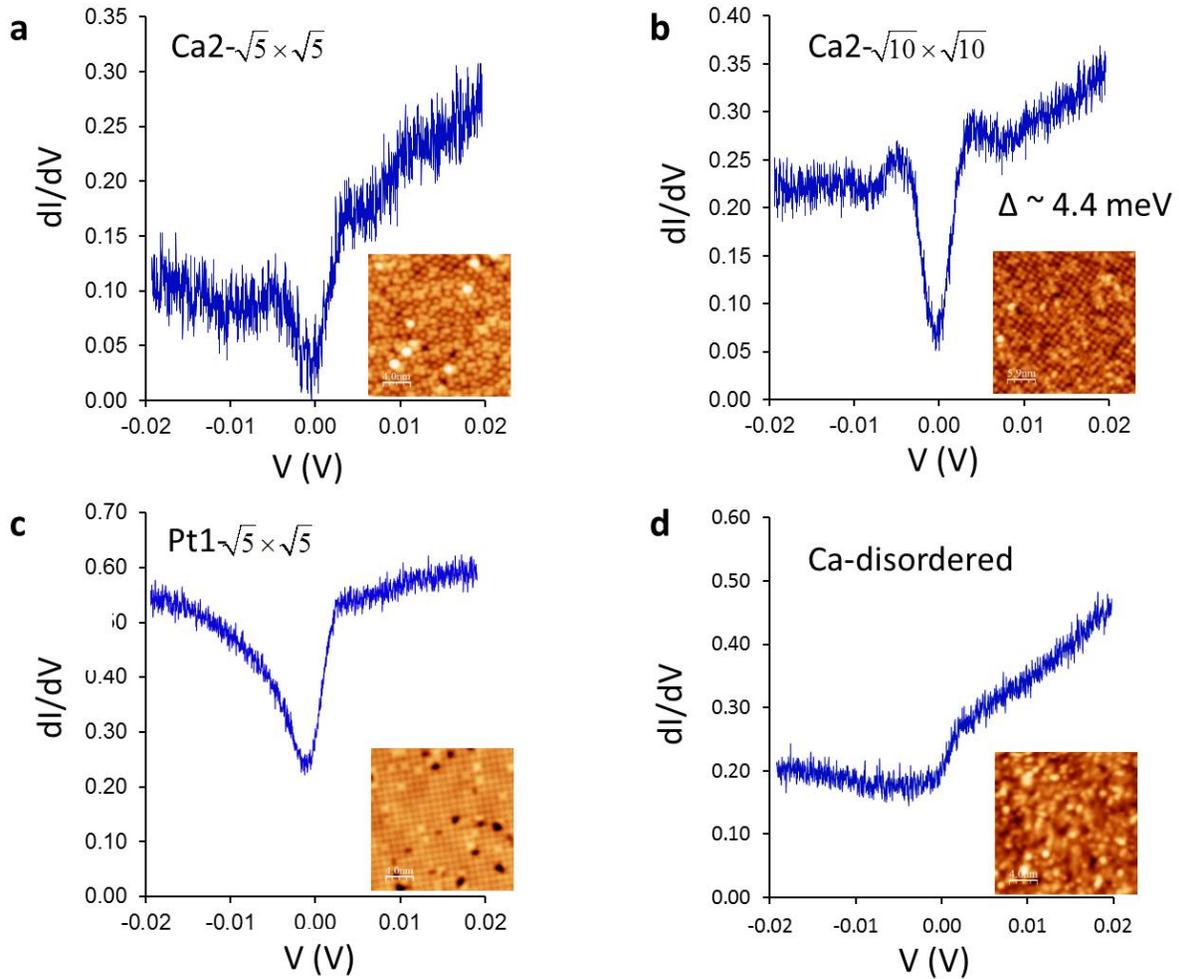

**Figure 5| STM images of four observed surfaces (inset) and spectra corresponding to each surface taken at 4.3 K:** (**a**) Ca2-$\sqrt{5}\times\sqrt{5}$, (**b**) Ca2-$\sqrt{10}\times\sqrt{10}$, (**c**) Pt1-$\sqrt{5}\times\sqrt{5}$, and (**d**) Ca-disordered. Only on the ordered Ca surfaces, superconducting features are observed. Note that spectra are raw data and the tip was stabilized at the sample bias $V_{sample}$ = 20 mV and the tunneling set current (**a**) 15 pA, (**b**) 20 pA, (**c**) 40 pA, and (**d**) 25 pA, respectively.

# Supplementary Information: Interrogating the superconductor $Ca_{10}(Pt_4As_8)(Fe_{2-x}Pt_xAs_2)_5$ Layer-by-layer

Jisun Kim[1], Hyoungdo Nam[2], Guorong Li[1], A. B. Karki[1], Zhen Wang[1,3], Yimei Zhu[3], Chih-Kang Shih[2], Jiandi Zhang[1], Rongying Jin[1], and E. W. Plummer[1,*]

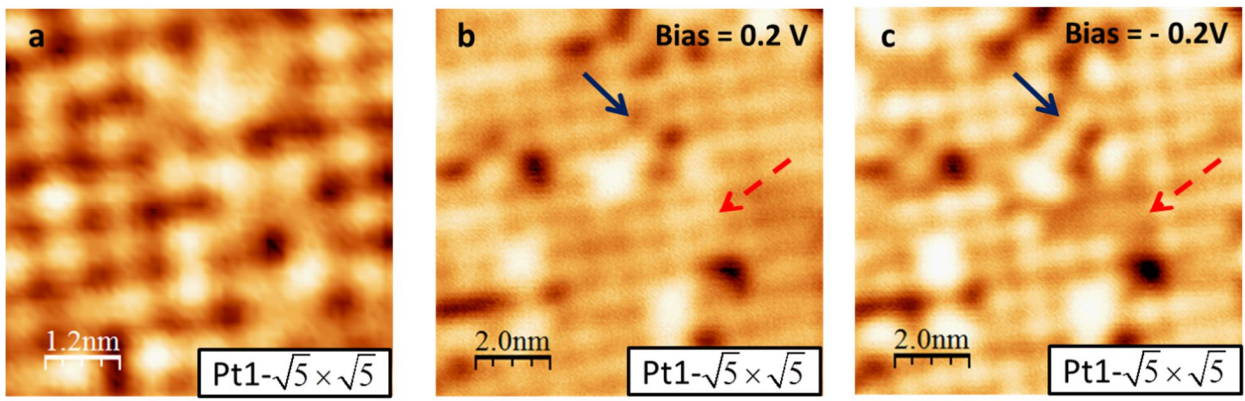

Supplementary Figure S1| **Local surface distortion of the Pt1-$\sqrt{5}\times\sqrt{5}$ surface:** (**a**) magnified image shows interconnection of certain Pt atoms which cannot be seen in larger images ($V_{sample}$ = 20 mV, I = 40 pA), (**b** – **c**) bias dependent surface structure (I = 2 pA). Certain Pt atoms (marked with broken red arrow) seem to be missing when imaged at a negative bias. The electronic inhomogeneity may be related to Pt deficiency in the $Pt_4As_8$ layer reported previously.[12]

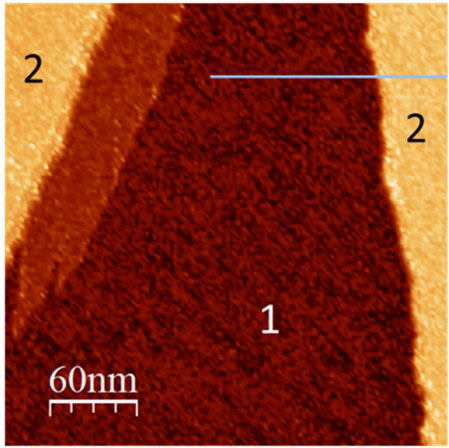
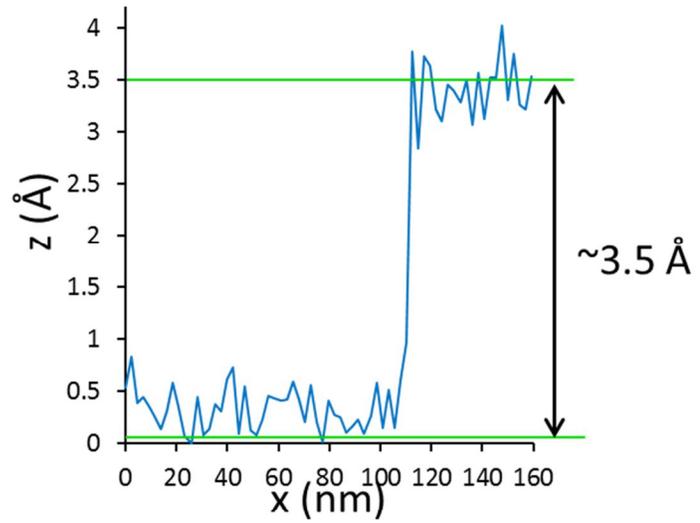
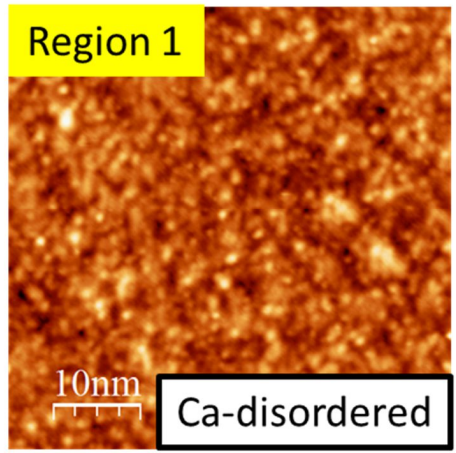
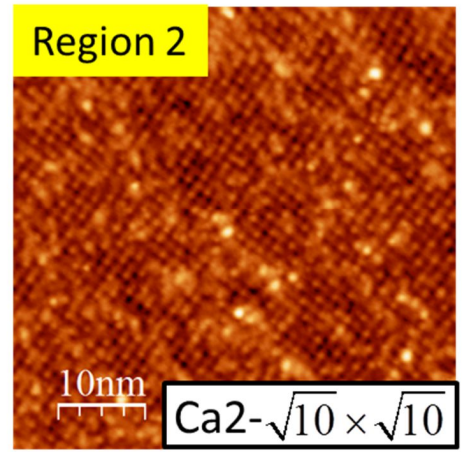

Supplementary Figure S2| **Ca-disordered surface and Ca2-$\sqrt{10}\times\sqrt{10}$ surface:** STM images of two different layers (the regions 1 and 2) after warmed from 4.3 K to room temperature (~ 290 K). The region 1 is disordered but the region 2 shows Ca2-$\sqrt{10}\times\sqrt{10}$ structure (the intermediate layer is due to the double-tip effect). The distance between the region 1 and 2 is about 3.5 Å, indicating that the region 1 is disordered Ca surface on the top of $Fe_2As_2$ layer. All images were acquired at $V_{sample}$ = - 0.5 V and I = 2 pA.

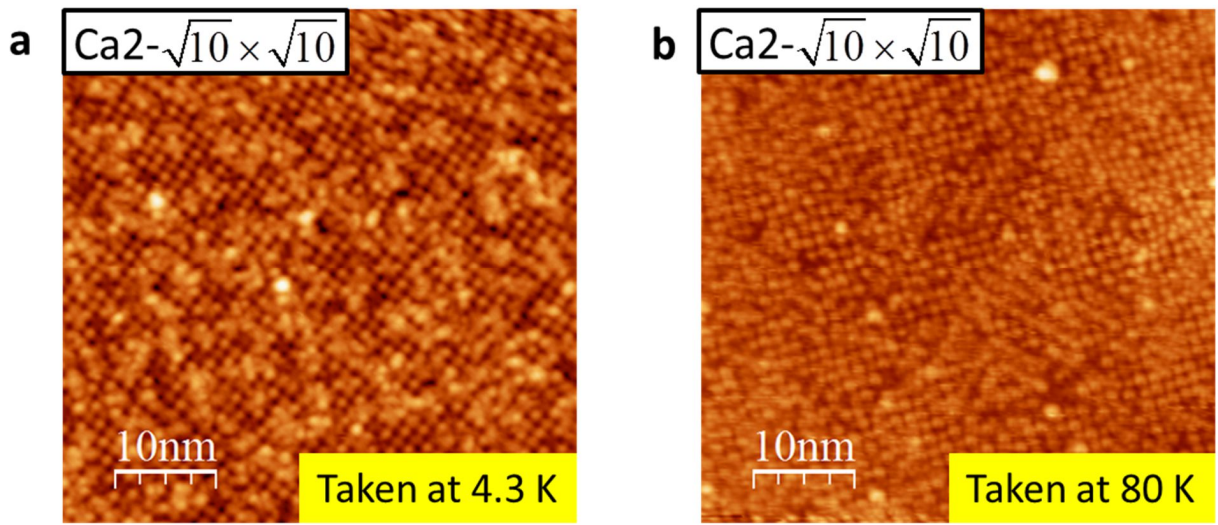

Supplementary Figure S3| **Ca2-$\sqrt{10} \times \sqrt{10}$ surface scanned at 4.3 K and 80 K:** STM image is taken at (a) 4.3 K, $V_{sample}$ = - 0.5 V, and I = 2 pA; (b) 80 K, $V_{sample}$ = 1 V, and I = 100 pA. There is no significant difference between (a) and (b).

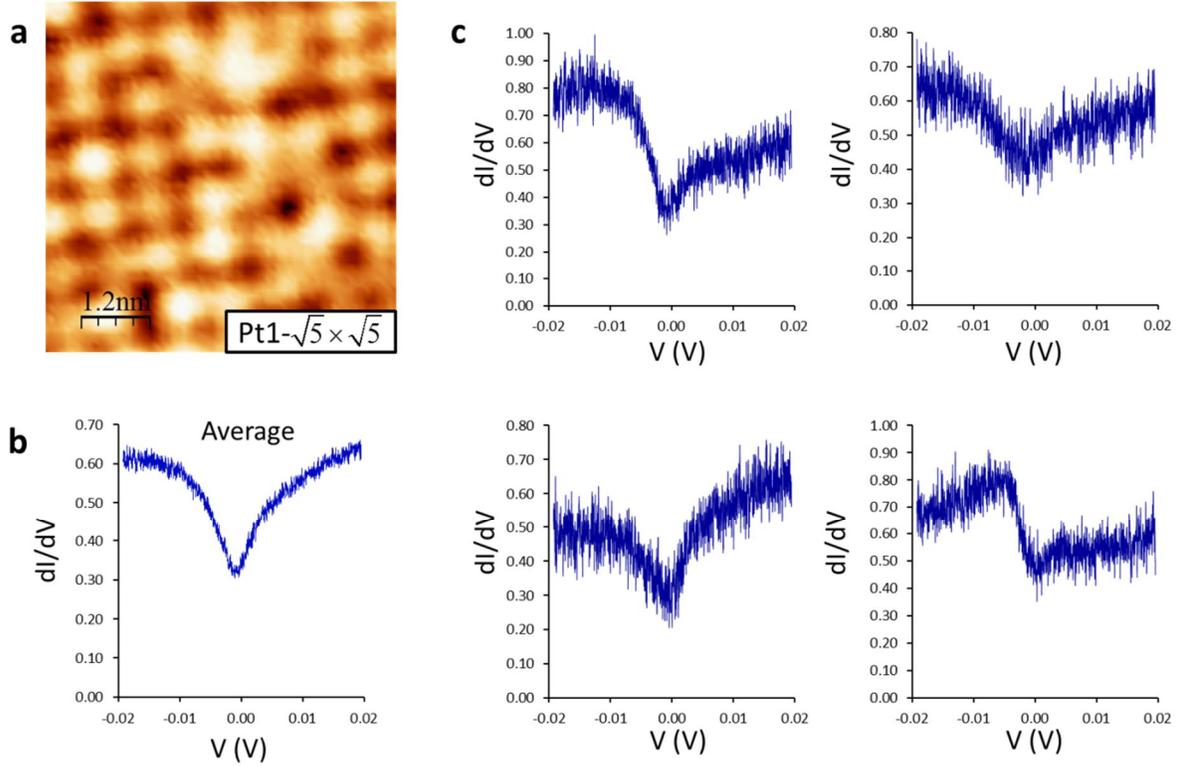

Supplementary Figure S4| **Site-dependent Pt1-$\sqrt{5}\times\sqrt{5}$ spectra:** (a) STM image of Pt1-$\sqrt{5}\times\sqrt{5}$ surface taken at 4.3 K ($V_{sample}$ = 20 mV, I = 40 pA); (b) 16×16 spectra average taken on the shown surface; (c) individual spectrum at different locations in (a). The individual spectrum clearly shows that the detailed feature (e.g. shape and ZBC) varies significantly depending on the site where STS is taken, even though the surface is well ordered. None of site-dependent spectra shows coherence peaks, indicating the Pt1-$\sqrt{5}\times\sqrt{5}$ surface is not superconducting.